\PassOptionsToPackage{pdfa}{hyperref}
\documentclass[sigconf, nonacm]{acmart}

\newcommand\vldbdoi{XX.XXXXX/XXXXXXX.XXXXXXX}
\newcommand\vldbpages{XXXX-XXXX}
\newcommand\vldbvolume{XX}
\newcommand\vldbissue{XX}
\newcommand\vldbyear{XXXX}
\newcommand\vldbauthors{\authors}
\newcommand\vldbtitle{\shorttitle}
\newcommand\vldbavailabilityurl{}
\newcommand\vldbpagestyle{empty}

\usepackage{booktabs}
\usepackage{subcaption}
\usepackage{color}
\usepackage{titlesec}

\newcommand{\system}{\emph{CAT}}

\titlespacing{\paragraph}{0pt}{3pt}{3pt}

\begin{document}

\title[Demonstrating CAT: Synthesizing Data-Aware Conversational Agents for Transactional Databases]{Demonstrating CAT: Synthesizing Data-Aware\\Conversational Agents for Transactional Databases}

\author{Marius Gassen}
\affiliation{\institution{Technical University of Darmstadt}}
\author{Benjamin Hättasch}
\affiliation{\institution{Technical University of Darmstadt}}
\orcid{0000-0001-8949-3611}
\author{Benjamin Hilprecht}
\affiliation{\institution{Technical University of Darmstadt}}
\author{Nadja Geisler}
\orcid{0000-0002-5245-6718}
\affiliation{\institution{Technical University of Darmstadt}}
\author{Alexander Fraser}
\affiliation{\institution{LMU Munich}}
\author{Carsten Binnig}
\orcid{0000-0002-2744-7836}
\affiliation{\institution{Technical University of Darmstadt}}

\begin{abstract}
Databases for OLTP  are often the backbone for applications such as
hotel room or cinema ticket booking applications.
However, 
developing
a conversational agent (i.e., a
chatbot-like
interface) to allow end-users to interact with an application using natural language
requires both immense amounts of training data and NLP expertise.
This motivates \system{}, which can be used to easily create conversational agents for transactional databases. 
The main idea is that, for a given OLTP database, \system{}
uses weak supervision to synthesize
the required training data
to train a state-of-the-art conversational agent,
allowing users to interact with the OLTP database.
Furthermore, \system{} provides an out-of-the-box integration of the resulting agent with the database.
As a major difference to existing conversational agents, agents synthesized by \system{} are
data-aware. This means that the agent decides which information should be requested from the user based on the current data distributions in the database, which typically results in
markedly
more efficient dialogues compared
with
non-data-aware agents.
We publish the code for \system{} as open source.

\end{abstract}

\maketitle

\pagestyle{\vldbpagestyle}
\begingroup\small\noindent\raggedright\textbf{PVLDB Reference Format:}\\
\vldbauthors. \vldbtitle. PVLDB, \vldbvolume(\vldbissue): \vldbpages, \vldbyear.\\
\href{https://doi.org/\vldbdoi}{doi:\vldbdoi}
\endgroup
\begingroup
\renewcommand\thefootnote{}\footnote{\noindent
This work is licensed under the Creative Commons BY-NC-ND 4.0 International License. Visit \url{https://creativecommons.org/licenses/by-nc-nd/4.0/} to view a copy of this license. For any use beyond those covered by this license, obtain permission by emailing \href{mailto:info@vldb.org}{info@vldb.org}. Copyright is held by the owner/author(s). Publication rights licensed to the VLDB Endowment. \\
\raggedright Proceedings of the VLDB Endowment, Vol. \vldbvolume, No. \vldbissue\ %
ISSN 2150-8097. \\
\href{https://doi.org/\vldbdoi}{doi:\vldbdoi} \\
}\addtocounter{footnote}{-1}\endgroup

\ifdefempty{\vldbavailabilityurl}{}{
\vspace{.3cm}
\begingroup\small\noindent\raggedright\textbf{PVLDB Artifact Availability:}\\
The source code, data, and/or other artifacts have been made available at \url{\vldbavailabilityurl}.
\endgroup
}

\section{Introduction}

\begin{figure}
\centering
\begin{subfigure}{0.48\columnwidth}
\includegraphics[width=\columnwidth]{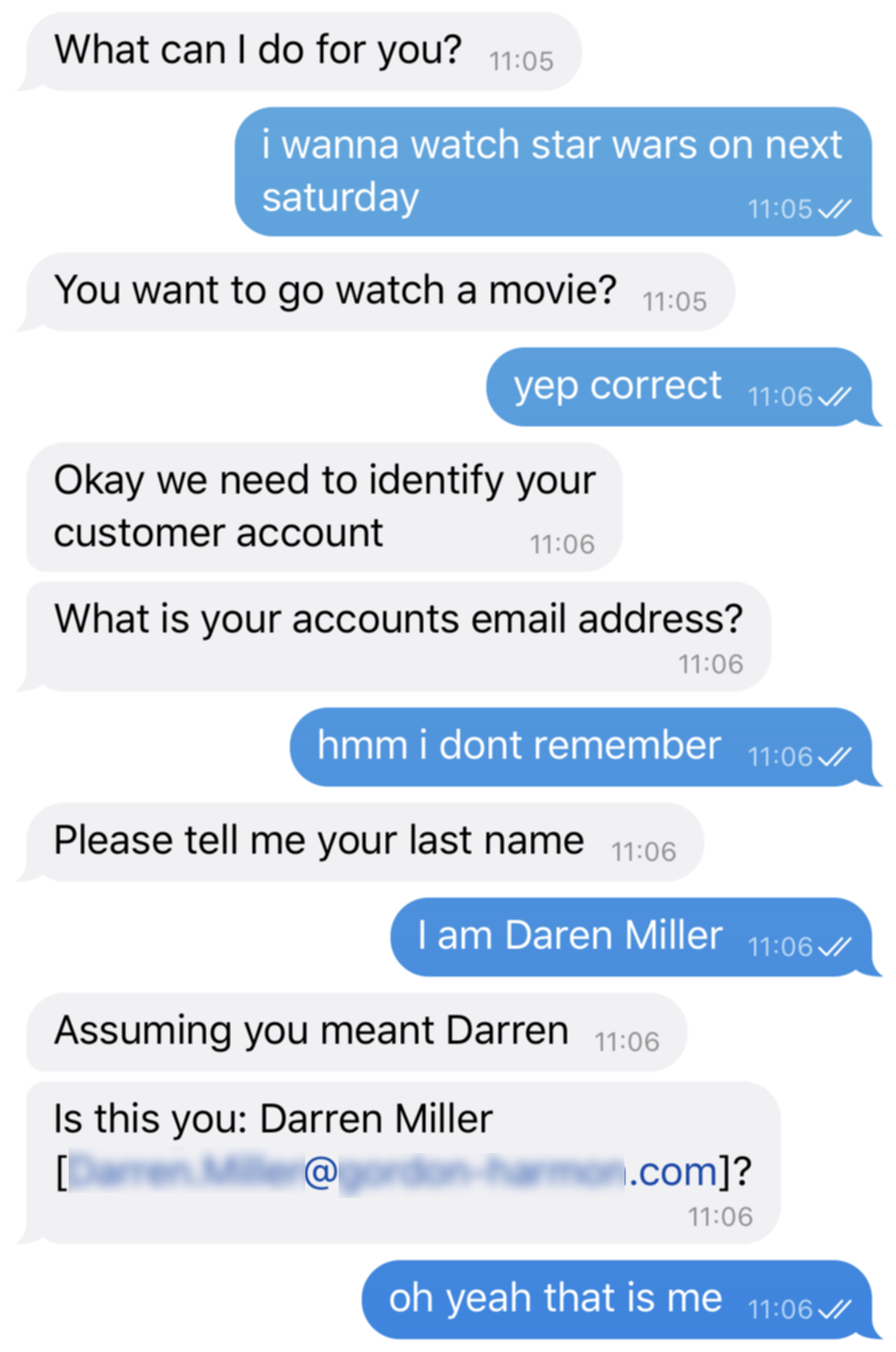}
\end{subfigure}
\begin{subfigure}{0.48\columnwidth}
\includegraphics[width=\columnwidth]{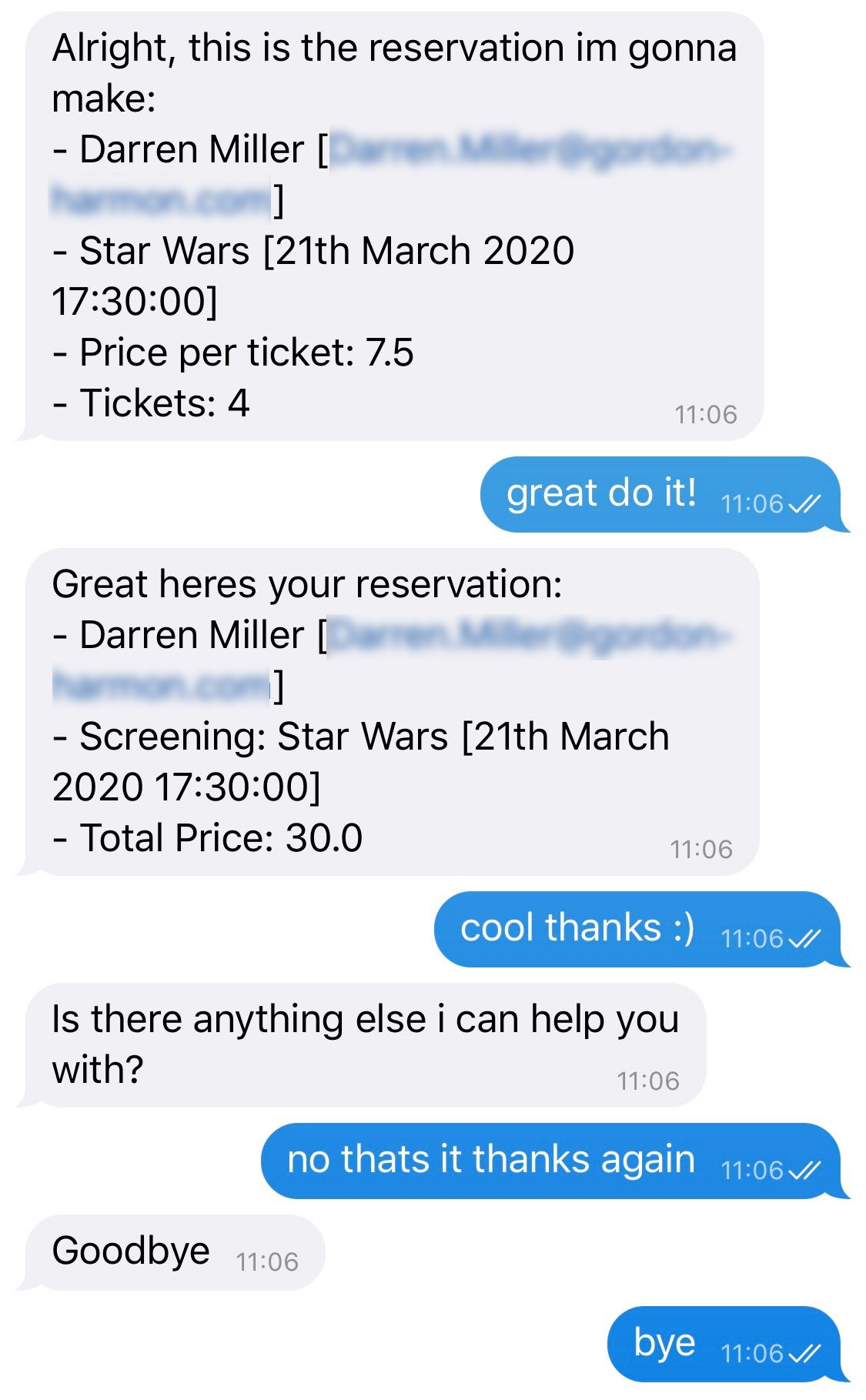}
\end{subfigure}	
\vspace{-2.5ex}
\caption{Exemplary Dialogue with a Conversational Agent synthesized by \system{}}
\vspace{-2.5ex}
\label{fig:demo}
\end{figure}

\paragraph*{Motivation} 

Natural language interfaces are becoming ubiquitous
because
they provide an intuitive way to interact with applications such as web shops, online ticketing systems,
etc. 
In particular, they allow users
to directly express their needs instead of having to remember application-specific commands or the correct usage of user interfaces. 
Moreover, consumer products like Amazon Alexa or Apple Siri further raise the
expectations
of customers to interact using natural language.
As a result,
companies
began
developing conversational agents for supporting simple tasks or even basic business processes. 
For instance, a customer of an insurance company could report a claim or check the status of an existing report using such a conversational agent.

Yet, developing a task-oriented dialogue system for a given OLTP application (e.g., allowing users to buy a movie ticket) is a daunting task
because
this not only requires large amounts of annotated training data (i.e., actual dialogues between users and the system) for every application but also a manual integration with the existing database. %

For instance, creating a conversational agent for a cinema ticketing system requires training data consisting of \emph{user utterances} (e.g., ``I want to reserve four seats tonight''), along with filled slots (e.g., \texttt{no\_seats}=4) and annotated \emph{user intents} (e.g., ``reserve seats'' or ``inform about available shows''). These dialogues, however, are expensive to gather and annotating them is
a large
manual
error-prone effort
which
requires
extensive
domain-knowledge. 
Worse,
neither the training dialogues nor the integration with the existing database can be
reused
for a different domain.

Another drawback of existing approaches to build task-oriented dialogue systems is the lack of integration between the  task-or\-iented dialogue system and the OLTP database, which is often the backbone of the business process. In current systems, a
large amount
of information must be provided manually even though it is already implicitly available in the database (for instance the required slots/attributes, the associated data types, the affected tables, etc.). 
Moreover, existing dialogue systems learn the order and types of information to request from the user purely from the manually created user dialogues.
Not taking the data characteristics into account
results in inefficient dialogues as we describe below. 

\begin{figure}
	\centering
	\includegraphics[width=0.95\columnwidth]{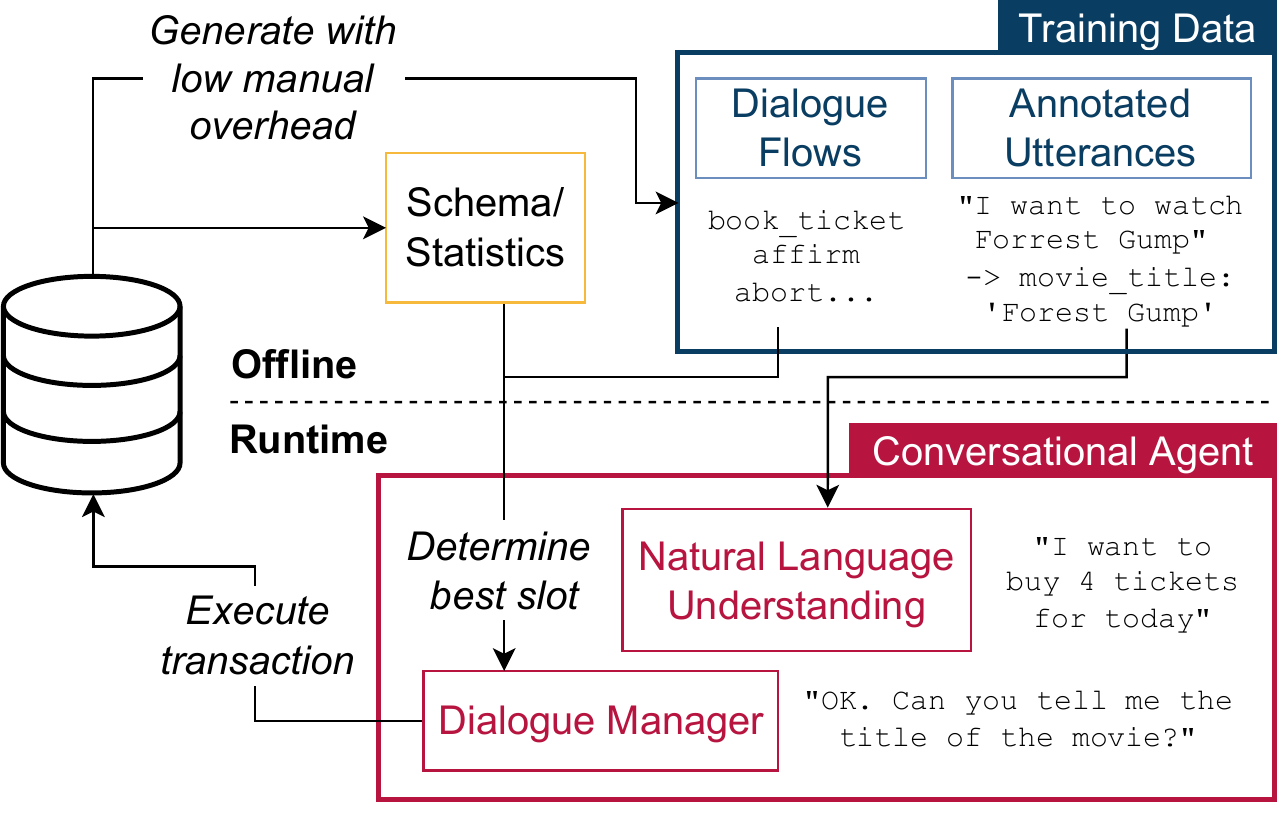}
	\vspace{-1.5ex}
	\caption{Overview of \system{}, showing both the creation and the usage of an agent.}
	\vspace{-4.5ex}
	\label{fig:overview}
\end{figure}

\paragraph*{Contributions} 
In this demo we introduce \system{}, a framework to synthesize conversational agents for a given database and a set of transactions (i.e., an OLTP workload with user-defined functions) with only minimal manual overhead. 
Given a database and a set of transactions, the user only has to provide a few example formulations for each intent instead of
a large number
of annotated example dialogues.
Using a data-driven simulation, our approach generates annotated dialogues of possible user interactions from those intents, which can then be leveraged to train a conversational agent. 
This alleviates the extensive process of manually creating dialogues, which has to be repeated for every domain and database. %

An inherent challenge is that for database transactions, it is often required to uniquely identify entities of the database. For instance, in order to book cinema tickets, the corresponding \emph{customer ID} is required. Often the customer will not have the unique ID at hand but only information such as their name or address. 
In contrast to existing conversational approaches, \system{} is data-aware; i.e., it considers the data characteristics at runtime to (1) deal with incomplete information (e.g., a customer who cannot remember an ID) and (2) request the most suitable information to narrow down the set of candidates as quickly as possible. 
Different from existing conversational approaches which take a pure learning-based approach to determine what to ask for, \system{} uses information such as database statistics (e.g., selectivities).
For example, once the user provided their name, the agent might ask
them for the city they live in, knowing that based on the entries in the database this is  sufficient to uniquely identify the \emph{customer ID} (while another name requires a different attribute to narrow down the options).

The
contributions
of this demo paper can be summarized as follows:
\vspace*{-0.5em}
\begin{itemize}
	\item\textbf{Automated Training Data Generation:} We suggest a procedure to automatically generate training dialogues given a database and a set of transactions with only minimal manual overhead. We then use it to train a conversational agent.
	\item\textbf{Data-driven Dialogue Policy:} We introduce a conversational agent policy that leverages the data characteristics to request information from the user to minimize the number of dialogue turns, i.e., to fulfill a user request as quickly as possible.
	\item\textbf{Demo Scenario:} We showcase \system{} by a demonstration scenario with a fully synthesized conversational agent for a movie database which allows a user to reserve tickets, cancel existing reservations and list movie theater screenings. We show both the creation of the agent using our system and the usage of the agent.
\end{itemize}
\vspace*{-0.5em}

\paragraph*{Outline} 
In the remainder of this paper, we first introduce the system architecture of \system{} (Section~\ref{sec:overview}), before we define our training data generation (Section~\ref{sec:training_data_generation}) and the data-driven dialogue policy (Section~\ref{sec:slot_selection}).
Finally,
we describe the demo application
(Section~\ref{sec:demo}).

\section{Overview of \system{}}
\label{sec:overview}

The goal of \system{} is to synthesize conversational natural language interfaces for database transactions while avoiding the shortcomings of existing task-oriented dialogue systems. %
To address these problems, \system{} leverages the information about a given database and a set of transactions: 
this is done for training data generation with weak supervision, but also at runtime to take data characteristics into account to steer the user dialogue (e.g., the movie a user wants to see) more efficiently.

For instance, a cinema could have a customer database storing the reservations for movie screenings. A typical transaction to make accessible using a conversational agent is the ticket booking process, where the users have to specify their \texttt{customer\_id}, the \texttt{screening\_id} and the number of tickets. In order to integrate such a task into a typical existing task-oriented dialogue system, we would first have to model the tasks the conversational agent supports (e.g., buy a ticket) along with slots, i.e., the required attributes for the task (e.g., the \texttt{screening\_id} and \texttt{customer\_id}).

All this information, however, is typically already available in the given database and the set of its transactions (e.g., implemented as stored procedures or user-defined functions). 
Therefore, the main idea of \system{} is to automatically extract and leverage this information instead of asking the user to manually specify it.
Moreover, \system{} then uses this information to synthesize annotated dialogues which are needed to train the conversational agent. Hence, instead of collecting this training data for every domain and database manually, we automate this process.
Moreover, the agent and the database are tightly integrated afterwards, and the agent can directly execute the desired transactions without any manual overhead---in contrast to
existing task-oriented dialogue systems where a dedicated database integration would have to be developed for every domain.

This tight integration also allows us to use characteristics of the given database (e.g., data-statistics) at runtime to guide the dialogue.
For instance, to identify
the
movie a user is interested
in,
the agent
asks
the users for properties of the movie (e.g., genre or actors playing in the movie).
In the following, we give a brief overview of how \system{} works as depicted in Figure~\ref{fig:overview}:

\paragraph*{Training Data Generation (Offline)} In order to generate a conversational agent, we require training data for both the natural language understanding (NLU) and the dialogue management (DM) models \cite{zhao-eskenazi-2016-towards}. The NLU model translates user utterances (e.g., "I want to watch 'Forrest Gump'") into annotated slots (\texttt{movie\_title='Forrest Gump'}) and user intents (ticket reservation). For the NLU training, we generate utterances using a few base templates that
are
provided by the developer.
To form full sentences from these templates, the existing data in the database can be used.
In addition, we increase the variety of the natural language by using automated paraphrasing, as done by Weir et al. \cite{weir2019dbpal} for natural language interfaces for databases. 
Furthermore, to learn typical dialogue flows, i.e., what high-level action to take next (e.g., retry a task after an abort), we generate additional training data using the idea of dialogue self-play \cite{DBLP:journals/corr/abs-1801-04871}, i.e., we simulate different users interacting with a conversational agent.
\system{} then uses this training data to train state-of-the-art models for NLU and DM using the RASA open source conversational AI framework.$\!$\footnote{\url{https://rasa.com/}}

\begin{figure}
    \setlength{\fboxsep}{.8em}
    \noindent\fbox{%
    \parbox{.92\linewidth}{%
    \vspace*{-.5em}
    \begin{footnotesize}
    \begin{center}
        \textbf{Database and Transaction}
        \vspace*{-.5em}
        \includegraphics[width=.95\columnwidth]{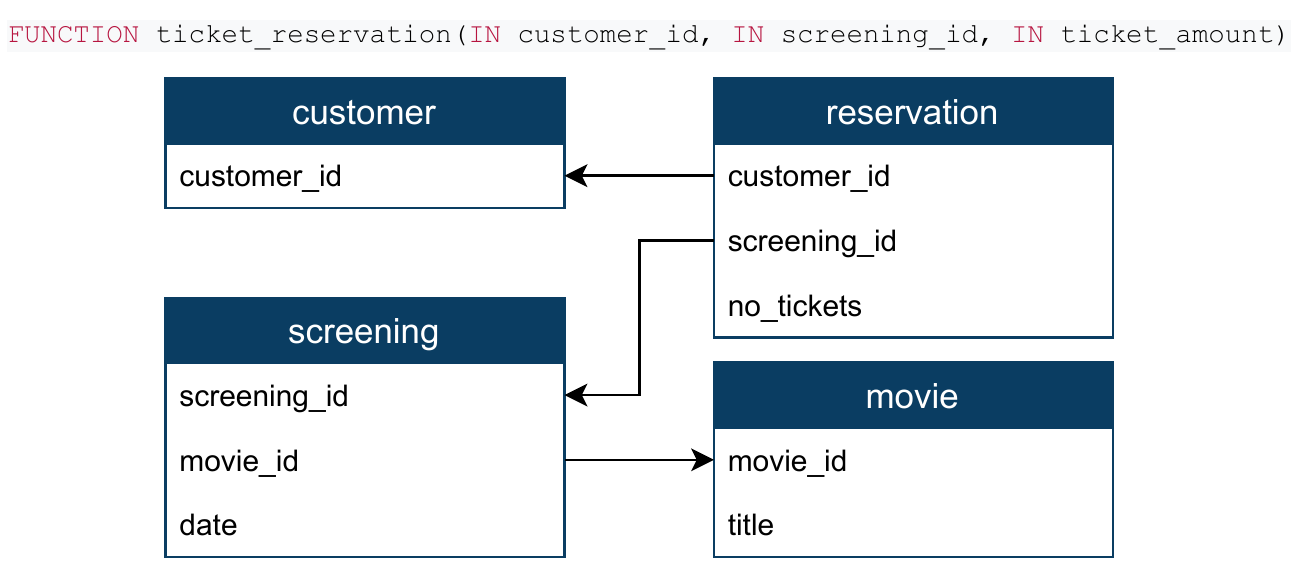}\\
        \textbf{Manually defined templates}
    \end{center}
			\textbf{Extracted Tasks and Schema Information}\\
			\texttt{ticket\_reservation}: \texttt{customer\_id} (customer), \texttt{screening\_id} (screening), \texttt{ticket\_amount} (integer), \dots
			
			\textbf{Natural Language Templates}\\
			The movie title is \{title\}; 
			I need \{no\_tickets\} tickets; 
			The screening is on the \{date\};
			\dots
		\begin{center}
            \textbf{Generated Training Data}
        \end{center}
        	\textbf{DM Training Data}\\
			\texttt{customer:} \texttt{request\_reservation}\\
			\texttt{bot:} \texttt{identify\_screening}\\
			\texttt{customer:} \texttt{abort\_task}
			
			\textbf{NLU Training Data}\\
			"The movie title is Forrest Gump." -> \texttt{intent:} \texttt{inform(movie\_title)}; \texttt{slots:}~\texttt{movie\_title='Forrest Gump'}
    \end{footnotesize}
    }}
    
    \caption{Exemplary inputs \& results for \system{}'s training data generation pipeline.}
    \vspace{-2ex}
    \label{fig:training_data}
\end{figure}

\paragraph*{Data-aware Dialogues (Runtime)} At runtime, the dialogue outlines created in the last step already determine the high-level flow of the dialogue.
In addition, the conversational agent has to decide on the low-level flow, to determine which information should be requested next from a user to uniquely identify an entity required for a task, e.g., it could decide to ask for the movie title to identify the movie.
In current approaches, this selection is usually done by learned models operating just on the previous input by this user \cite{DBLP:journals/corr/abs-1801-04871}.
In contrast, in order to efficiently narrow down the candidate movies, \system{} takes information such as the selectivity of attributes already in the database into account. In addition, we allow adding an annotation to the database schema
indicating which of the
attributes are probably unknown to the customer. For instance, even though the \texttt{screening\_id} is very useful and ultimately required for the transaction, the user will most likely not be aware of it and the conversational agent should thus not request it from the user. This results in more succinct dialogues, since the agent quickly gathers the information needed for a transaction.

In particular, the best information (i.e., a so-called slot) to request depends on (i) the probability that the user knows a certain attribute and (ii) how much this attribute narrows down the current set of candidates. Learning both factors end-to-end means learning the database content along with user preferences simultaneously, and again requires
a large
amount of data. We thus propose a different approach and explicitly keep track of the candidates (e.g., the screenings that match the previous user preferences) and request the next attribute based on the data distribution of the candidates and the likelihood that the user can provide this information. 
Moreover, while existing task-oriented dialogue systems implicitly assume that the database consists of just a single table \cite{DBLP:journals/corr/abs-1801-04871}, we can seamlessly integrate foreign-key dependencies, e.g., a user can provide information about actors to narrow down the set of possible screenings via the movie relation. 

Another advantage of this data-awareness is that no retraining is required in case data changes. The updated database is simply leveraged at runtime to steer the dialogue. %

\section{Training Data Generation}
\label{sec:training_data_generation}

Both the natural language understanding (NLU) and dialogue management (DM) \cite{zhao-eskenazi-2016-towards} components are learned models and thus require dedicated training data, which is expensive to collect. Consequently, we try to automate the training data generation as much as possible. %
We now describe the training data generation pipeline for both models, examples for inputs and results can be found in Figure~\ref{fig:training_data}:

\paragraph*{Dialogue Management (DM)} %
The high-level flow of dialogues in \system{}
is
derived from
training data
synthesized using a so-called dialogue simulation \cite{DBLP:journals/corr/abs-1801-04871}.
\system{} simulates typical dialogues between the conversational agent and the user
who
communicate with each other using predefined actions (e.g., \texttt{request\_reservation}).
The set of possible actions in \system{} is derived automatically from the transaction definition.

By sampling different user behavior during the simulation (e.g., sometimes performing the whole action and sometimes aborting it) the synthesized dialogue flows consist of different outlines that are later incorporated into the agent. 
Different from Shah et al. \cite{DBLP:journals/corr/abs-1801-04871}, we do not model the process of uniquely identifying entities in detail in this dialogue self-play, e.g., asking for the right slots to find the exact screening is not incorporated in this step. 
Instead, we only include the high-level action (e.g., \texttt{identify\_screening}, see Figure~\ref{fig:training_data}). 
Which information from a set of candidates is requested to uniquely identify the screening is then decided at runtime (see Section \ref{sec:slot_selection}).

\paragraph*{Natural Language Understanding (NLU)}
Moreover, in addition to the training data for high-level dialogue flows, \system{} also synthesizes training data for the NLU model.
To this end, we require utterances of a user ('I want to see the movie 'Forrest Gump'") along with annotated slots (\texttt{title='Forrest Gump'}) and user intents (e.g., reserve a ticket or ask for information about a movie) as ground truth labels. Gathering this information is a substantial manual effort---collecting dialogues would come at the cost of simulating dialogues with testers. Even if dialogue traces are available, annotating them with the intents remains a manual effort. We thus take a different route, and let the developer specify a few natural language templates (e.g., "I want to watch \{\texttt{movie\_title}\}").
By filling the placeholders with actual data stored in the database, we synthesize annotated natural language statements, which we automatically paraphrase afterwards to further augment the training data. Different from Shah et al. \cite{DBLP:journals/corr/abs-1801-04871} where the user similarly specifies templates, we do not use crowdsourcing for this since this incurs high costs and might not be feasible for many transactions but instead
utilize
automated paraphrasing approaches.

\paragraph{Initial Evaluation Results}
We compared several configurations of \system{} to state-of-the-art approaches for intent classification and slot filling, using the widely used ATIS spoken conversation corpus \cite{atis}.
While all baselines require manually crafted training data, \system{}
only
relies on synthesized training data,
but still reaches
comparable performance for slot filling. Moreover,  on the intention classification task, \system{}
even
outperforms multiple
baselines.

\section{Data-Aware Dialogues}
\label{sec:slot_selection}

\begin{figure}
	\centering
	\includegraphics[width=0.9\columnwidth]{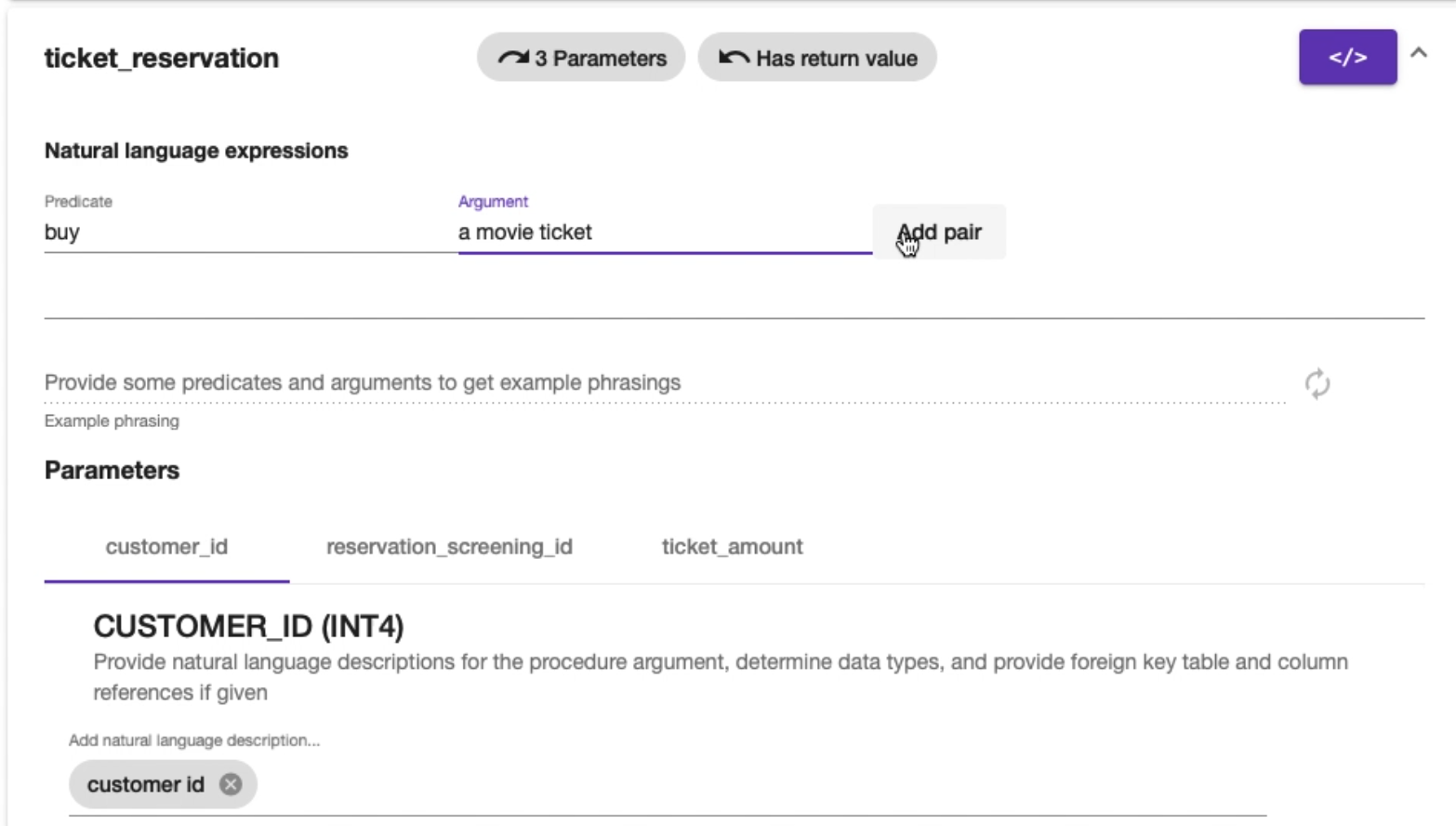}
	\vspace{-1.5ex}
	\caption{Schema Annotation in \system{}'s GUI}
	\vspace{-2.5ex}
	\label{fig:anno}
\end{figure}

We decide which information to request from the user for the unique identification of entities (e.g., ask for the movie title to find the screening) at runtime by keeping track of the current set of candidate entities (e.g., screenings that match with the already expressed user preferences) and select
those
attributes which narrow down this set as quickly as possible,
the
\emph{informative} attributes.
To do this, we
choose the attribute with the highest entropy. 

Note that the optimal attribute is not necessarily part of the table storing the entity. For instance,
if
a customer does not recall the exact movie title, it might be beneficial to ask for actors appearing in the movie. 
Since keeping track of candidates happens at runtime, it is not feasible to join every possible table with the set of candidates. 
Instead, we employ a priori information on the number of unique values of an attribute as well as the distribution of which attributes users were aware of in previous sessions, and iteratively join additional tables to the current candidate set to provide improved next attributes to request from the user. 

However, informative attributes are not useful if the user is not aware of them, e.g., while customer IDs quickly narrow down the set of customers, it is very unlikely that the user has such an ID at hand. Hence, the second dimension is the \emph{User Awareness}.
We address this two-fold: First, the developer can specify that certain attributes should preferably not be requested, e.g., IDs or other technical fields. Second, we learn from interactions with the conversational agent which attributes the users are likely to know. We combine both this probability and the informativeness of the attribute to score candidate attributes to request next.

\paragraph{Initial Evaluation Results}
To evaluate the effectiveness of our data-aware selection policy, we compared it to static and random selection strategies using a movie database and again the ATIS dataset.
The speedup (in terms of interaction turns) compared to a random strategy can be up to $80\,\%$ for large tables with many dimensions to join.
When large amounts of data similar to the production entries are already available at training time, the static strategy can reach a similar performance as our data-aware policy,
but will not adapt to data distribution changes at runtime.
Additionally, it cannot react to systematic problems in uniquely identifying entries of some tables (caused by data characteristics like almost identical entries).
An integrated caching strategy leads to an average response latency of only a few milliseconds.

\section{Demonstration Scenario}
\label{sec:demo}

In our demo, we showcase how a conversational agent for a cinema database supporting screening reservations and cancellations can be synthesized. It is fully integrated with the underlying database and allows users to interact using natural language to complete the domain-specific tasks. 

To synthesize the required training data, we first annotate the schema and provide several natural language templates for the transactions using \system{}'s GUI, as depicted in Figure~\ref{fig:anno}. 
This is in fact the only database-dependent task for developers who want to synthesize an agent.
We then start our training data generation to obtain both natural language statements for the NLU model and dialogue flows for the DM models. Afterwards, we trigger the training of these state-of-the-art models and generate the integration code with the database.
With the completion of these steps, we have synthesized a conversational agent which interacts with users and triggers the right database transaction with the correct parameters at runtime.

The users can use this trained conversational agent to interact with the database as depicted in Figure~\ref{fig:demo}. For instance, if the user wants to buy movie tickets, the agent will request the required information and execute the transaction upon confirmation.
In the demo video, it can be seen how the agent identifies the intents and reacts to the user statements. It uses the information entered to identify their account, corrects misspellings, and asks the user to choose from a list of screenings fulfilling the preferences they have expressed.
Finally, this triggers the execution of the transaction and the result is shown.

\bibliographystyle{abbrv}
\bibliography{main}

\end{document}